\documentclass[pra,preprint,nofootinbib,superscriptaddress,tightenlines]{revtex4}
\usepackage[english]{babel}
\usepackage[latin1]{inputenc}
\usepackage{graphicx}
\usepackage{amsmath}
\usepackage{amssymb}
\usepackage{color}
\usepackage{ulem}
\usepackage{bm}
\usepackage{xcolor}
\usepackage{hyperref}
\hypersetup{
    colorlinks,
    linkcolor={red!50!black},
    citecolor={blue!50!black},
    urlcolor={blue!80!black}
}

\usepackage{units}
\usepackage{subfigure}
\usepackage{enumitem}

\newcommand{\etal}{\textit{et~al.}}
\newcommand{\bs}[1]{\boldsymbol{#1}}

\newcommand{\ket}[1]{\left|#1\right\rangle}

\newcommand{\spec}[2]{{}^{#1}\!#2}
\newcommand{\Iso}[2]{{}^{#2}\text{#1}}
\newcommand{\Cl}[3]{\text{C}_{#1,#2}^{#3}}

\def\d{\mathrm{d}}
\newcommand{\idz}{i\partial_0}

\newcommand{\galder}{\overleftrightarrow{\bs{\nabla}}}
\newcommand{\kinterm}[1]{\frac{\bs{\nabla}^2}{#1}}
\newcommand{\tens}[2]{\left\lbrace #1\right\rbrace_{#2}}
\newcommand{\Hc}{\text{H.c.}}

\newcommand{\onepl}{{(1^+)}}
\newcommand{\aB}{a_\text{B}}
\newcommand{\ResFactor}{a_{1,\text{bg}}\,\Delta B}
\newcommand{\kB}{k_\text{B}}
\newcommand{\kT}{k_T}
\newcommand{\kres}{k_\text{res}}

\newcommand{\Li}{\Iso{Li}{6}}
\newcommand{\Lie}{\Iso{Li}{6}_2(\text{e})}
\newcommand{\Lid}{\Iso{Li}{6}_2(\text{d})}

\newcommand{\args}[3]{\left(#1,\,#2;\,#3\right)}
\newcommand{\momint}[2]{\int\!\frac{\text{d}^{#1}#2}{(2\pi)^{#1}}}

\begin{document}
\title{Three-body losses of a polarized Fermi gas near a $\bs{p}$-wave Feshbach resonance in effective field theory}
\author{M. Schmidt} \affiliation{Institut f\"ur Kernphysik, Technische Universit\"at Darmstadt, 64289 Darmstadt, Germany
}
\author{L. Platter} \affiliation{Department of Physics and Astronomy, University of Tennessee, Knoxville, TN 37996, USA}
\affiliation{Physics Division, Oak Ridge National Laboratory, Oak Ridge, TN 37831, USA}
\author{H.-W. Hammer} \affiliation{Institut f\"ur Kernphysik, Technische Universit\"at Darmstadt, 64289 Darmstadt, Germany
}
\affiliation{ExtreMe Matter Institute EMMI, GSI Helmholtzzentrum f\"ur Schwerionenforschung GmbH,
64291 Darmstadt, Germany
}
\date{\today}

\begin{abstract}
  We study three-body recombination of fully spin-polarized $^6$Li
  atoms that are interacting resonantly in relative
  $p$-waves. Motivated by a recent experiment, we focus on negative
  scattering volumes where three atoms recombine into a deep
  dimer and another atom. We calculate the three-body recombination rate
  using a Faddeev equation derived from effective field theory. In particular,
  we study the magnetic field and temperature dependences of
  the loss rate and use the recombination data to determine the
  effective range of the $p$-wave atom-atom interaction. We also
  predict the existence of a shallow three-body bound state that
  manifests itself as a prominent feature in the energy-dependent
  three-body recombination rate.
\end{abstract}
\preprint{INT-PUB-19-055}
\smallskip
\maketitle
\section{Introduction}
\label{Sec:Intro}
Experiments with ultracold atomic gases provide a unique way to
explore the interactions between atoms. Specifically, strongly
interacting system have recently received a lot of
attention~\cite{Braaten:2004rn,RevModPhys.80.1215,RevModPhys.82.1225,
Naidon:2016dpf,RevModPhys.89.035006}. For
example, the loss rate of an ultracold gas of strongly interacting
bosons will display discrete scale invariance when it is measured as a
function of the scattering length. This discrete scale invariance
is related to the well-known Efimov effect~\cite{Efimov:1970zz}
and leads to a log-periodic dependence of the 
three-body loss rate on the scattering
length~\cite{PhysRevLett.83.1566,PhysRevLett.83.1751,Bedaque:2000ft,Braaten:2001hf}. Three-body
losses occur in ultracold atomic gases as a result of three-body
collisions in which the atoms gain kinetic energy due to the formation
of a two-body bound state. Indeed, the Efimov effect was first observed through
its signature in three-body losses in a cold gas of Cs
atoms~\cite{Kraemer:2006nat}.

Identical fermions cannot interact in a relative $s$-wave due to the
Pauli principle. However, their $p$-wave scattering volume can be
tuned and a number of experiments have examined the
features of strongly interacting Fermi gases~\cite{PhysRevA.70.030702,PhysRevA.71.045601,PhysRevLett.90.053201,Yoshida:2018,Waseem:2018,Waseem:2019}.
The Efimov effect does
not occur in these
systems~\cite{Jona-Lasinio:2007,Nishida:2011np,Braaten:2011vf},
however, losses still occur through
recombination processes into shallow or deep dimers.
The three-body losses in spin-polarized ultracold gases of $^6$Li
have recently been studied in the group of Takashi Mukaiyama at
Osaka University focusing on scaling laws \cite{Yoshida:2018},
unitary-limited behavior \cite{Waseem:2018}, and the role of cascade processes
\cite{Waseem:2019}.
     
In this paper, we focus on the data taken by Waseem \etal\ in
\cite{Waseem:2018}.  Specifically, they considered the
$\ket{F=1/2,\,m_F=+1/2}$ hyperfine state and measured the loss rate at
large negative $p$-wave scattering volume, enhanced by a Feshbach
resonance at $B_0 = 159.17(5)$~G. At negative scattering volume, three
atoms recombine into a so-called deep dimer, {\it i.e.}, a dimer whose
binding energy is so large that it cannot be described by the
parameters of the effective range expansion.  The authors of this work
used a simplified Breit-Wigner model for the energy-dependent
three-body recombination rate coefficient $K_3(E)$
to determine information about the atom-atom interaction
and to model the temperature-dependent three-body loss coefficient
$L_3(T)$. While they managed to reproduce the measured loss
data appropriately, we will discuss in detail that their
simplified approach has also some important limitations.

It is particularly important to have an accurate microscopic
description for the three-body recombination if the goal is to extract
two-body observables from three-body processes with some understanding
of the resulting uncertainties.  Various approaches can be used to
develop a microscopic description of this process.
Effective field theory (EFT) uses the separation between short- and
long-range scales in a system to construct a controlled expansion. It
has been applied successfully in particle, nuclear and atomic
physics~\cite{Braaten:2004rn, Hammer:2010kp,Hammer:2019poc}.
The parameters that appear in the EFT description of atomic systems
with strong interactions can be related directly to the effective
range parameters.  This approach is therefore model independent and
facilitates an unbiased analysis of experimental data.
Systems with $p$-wave
interactions have been studied using the short-range effective field theory
previously~\cite{Braaten:2011vf}. It was
found that a real three-body parameter
is required for renormalization.
However, the emphasis of this work was on renormalization issues and
the spectrum of three-body states.

Here, we will use EFT to study the three-body loss rate into deep
dimers at finite temperature with the parameters relevant to the
experimental measurements by Waseem \etal~\cite{Waseem:2018}.
We will construct the interaction of two atoms in a relative $p$-wave
and use it to derive an integral equation whose solution allows us to
compute the three-body recombination rate $K_3(E)$. Temperature
averaging then yields the three-body loss coefficient $L_3(T)$.
Comparison with the data obtained by Waseem \etal\ will let us draw
conclusions about the features of the two-body interaction such as the
$p$-wave effective range. Moreover, it will allow us to tie these
features to other three-body observables such as the energy of a
shallow three-body bound state.

This manuscript is organized as follows. In
Secs.~\ref{sec:three-body-recomb} and \ref{sec:diat-atom-ampl}, we
will introduce our microscopic framework to calculate the three-body
loss coefficient. Some details of the EFT framework are given in the
Appendices. Our results are presented and compared to the data by
Waseem {\it et al.}~\cite{Waseem:2018} in
Sec.~\ref{Sec:Results}. We end with a short summary and outlook in
Sec.~\ref{sec:summary}.

\section{Recombination near a $p$-wave resonance}
\label{sec:three-body-recomb}
At sufficiently low energies, the elastic scattering properties of
particles can be quantified using the effective range expansion. In the
$p$-wave, the expansion is
\begin{eqnarray}
  \label{eq:ere}
  k^3 \cot\delta_1 = -\frac{1}{a_1} + \frac{r_1}{2} k^2+\ldots~,
\end{eqnarray}
where $\delta_1$ is the scattering phase shift,
$a_1$ denotes the scattering volume,
$r_1$ is the effective range, and the ellipses denote higher order
terms. We note that our definition of the effective range
$r_1$ differs by a factor of
$-2$ from the definition used in Ref.~\cite{Waseem:2018}.

\begin{figure}
        \includegraphics[scale=1]{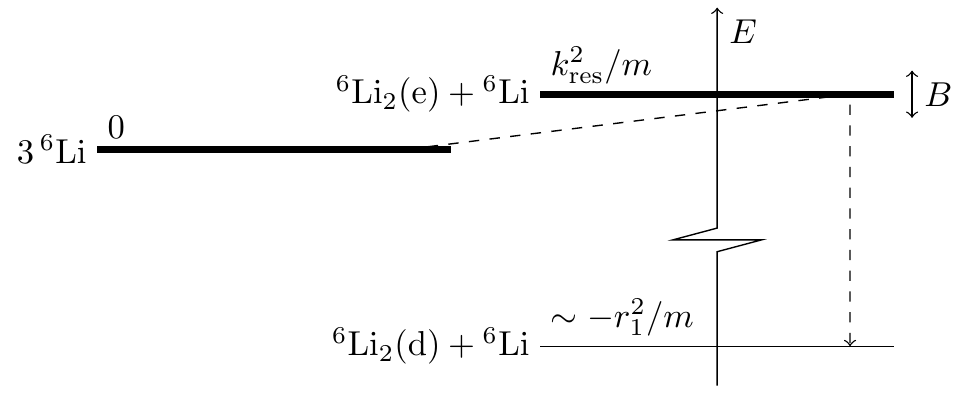}
        \vspace{-0.5cm}
	\caption{\label{Fig:Levels} Level scheme of three
          spin-polarized $\Li$ atoms in the hyperfine state
          $\ket{F=1/2,\,m_F=+1/2}$. Three-body recombination into a
          deeply bound state $\Lid$ proceeds through the Feshbach
          resonance state $\Lie$. Its position
          $\kres^2(B)/m$ can be tuned by a magnetic field
          $B$.}
\end{figure}

The level scheme of three spin-polarized $\Li$ atoms in the hyperfine
state $\ket{F=1/2,\,m_F=+1/2}$ is illustrated in
Fig.~\ref{Fig:Levels}.  Two identical $\Li$ atoms form a deeply bound
state denoted $\Lid$.  The Feshbach resonance creates an excited state
$\Lie$, whose position can be tuned by changing the magnetic field
$B$.
The scattering volume of two $\Li$ atoms in the
$\ket{F=1/2,\,m_F=+1/2}$ state, $a_1$, diverges at the resonance 
position, $B_0= \unit[159.17(5)]{G}$. For $B\approx B_0$, the magnetic field
dependence has the form
\begin{equation}
	\label{Eq:Feshbach}
	a_1(B)= a_{1,\text{bg}}\left(
		1+\frac{\Delta B}{B-B_0}
	\right)\approx \frac{\ResFactor}{B-B_0},
\end{equation}
where $a_{1,\text{bg}}<0$ is the background scattering volume and
$\Delta B>0$ is the resonance width \cite{Waseem:2018}.\footnote{
  For detunings $B-B_0<\unit[0.5]{G}$ as in \cite{Waseem:2018},
  the constant term in Eq.~\eqref{Eq:Feshbach} is less than $1\%$ of
  the total and can be neglected.}
As input, we use the value
$\ResFactor=\unit[-2.8(3)\times 10^6\,\aB^3]{G}$, 
obtained in a fit to the thermalization rate of the spin-polarized
$\Li$ gas by Nakasuji \etal~\cite{PhysRevA.88.012710}.
For fixed $B\approx B_0$,
$a_1(B)$ is then given with an uncertainty of roughly $\unit[30]{\%}$.

The $p$-wave effective range $r_1$ is usually assumed to depend weakly
on $B$ in the immediate vicinity of $B_0$~\cite{PhysRevA.78.063616,
  PhysRevA.88.012710, Waseem:2018}. Waseem~\etal\ suggested the
near-resonance estimate
\begin{equation}
	\label{Eq:FermireValue}
	(r_1)_\text{est}\equiv \frac{2}{m\ \delta\mu\ \ResFactor}
	=-0.182(20)\,\aB^{-1}\,,
\end{equation}
and used it in their analysis of the experimental
data~\cite{Waseem:2018}. In Eq.~\eqref{Eq:FermireValue},
$\delta\mu=\unit[113(7)\,\kB]{\mu K\,G^{-1}}$ denotes the relative
magnetic moment between $\Lie$ and two $\Li$ atoms with $\kB$ being Boltzmann's
constant. We adopt the assumption that $r_1$ is constant in
$B$. However, we note that different estimates for $r_1$ have been
given that differ significantly from $(r_1)_\text{est}$. First,
Bruun~\etal\ derived Eq.~\eqref{Eq:FermireValue} only for two bosons
near an $s$-wave Feshbach resonance~\cite{PhysRevA.71.052713}. Second,
Nakasuji~\etal\ obtained a different value $-0.116(10)\,\aB^{-1}$ in
their fit to the thermalization rate~\cite{PhysRevA.88.012710}. They also
cited an even smaller theory prediction $-0.096(6)\,\aB^{-1}$ by
Julienne (Ref.~[29] of their work). This value deviates by roughly
$\unit[50]{\%}$ from Eq.~\eqref{Eq:FermireValue}. Thus, $r_1$
introduces the largest uncertainty to the study.
It is one goal of this work to predict $r_1$ from data of the
three-body loss coefficient $L_3$.

In the experiment by Waseem~\etal, recombination was studied for
$B-B_0>0$ ($a_1<0$), where the process can be distinguished from
background losses~\cite{Waseem:2018}. Thus, we restrict
ourselves to this region when calculating $L_3$. On this side of the
Feshbach resonance, the two-body system has a resonance pole above
threshold representing the $\Lie$ state.\footnote{For $B-B_0<0$
  (corresponding to $a_1>0$), $\Lie$ is a shallow bound state.} The
position of the corresponding maximum of the scattering amplitude on
the real $k$ axis will be denoted by $\kres$
(cf. Fig.~\ref{Fig:Levels}). It varies with $B$ and will be denoted
``resonance momentum'' in the following. For large scattering volume,
the resonance momentum can be approximated by \cite{Bedaque:2003wa,tuprints8778}
\begin{equation}
\kres(B) = \sqrt{\frac{2}{a_1(B) r_1}}\;.
\label{eq:defkres}
\end{equation}
Note that the Feshbach resonance introduces a strong
  separation of momentum scales to the two-body system. In particular,
  the resonance momentum $\kres(B)$ is much smaller than the natural
  (high) momentum scale set by the effective range $r_1$. Such a
  separation is an important requirement for the EFT approach. In the
  system at hand, it enables an expansion of the two-body scattering
  amplitude in terms of the ratio $\chi_2(B)\equiv \kres(B)/r_1\ll
  1$. This expansion yields a simple Breit-Wigner-like diatom
  propagator at leading order in the expansion in $\chi_2(B)$
  \cite{tuprints8778}. In the following, we restrict our analysis to
  leading order (LO) in the expansion in $\chi_2(B)$.

For $B-B_0>0$, three-body recombination proceeds in the absence of a
shallow dimer state only into deep dimer states $\Lid+\Li$. Such a process
involves a large excess of kinetic energy $\sim {r_1}^2/m$ outside our
EFT's applicability region, {\it i.e.}, the recombination process
happens when all three atoms are very close together. While the process cannot
be described in detail in the framework of our EFT, the total rate for
recombination into deep dimers can be described by making the
three-body parameter complex~\cite{Braaten:2003yc}. This corresponds
to using an optical three-body potential to model the losses at
short distances which can be treated in perturbation
theory~\cite{Braaten:2001hf}.

As shown by Esry \etal, the recombination rate of three identical
fermions vanishes at total kinetic energy $E=0$
\cite{PhysRevA.65.010705}. More specifically, it obeys the threshold
law $K_3\propto E^2$ in the partial wave channel $J^P=1^+$ and is suppressed by
further powers of $E$ for other $J^P$. Thus, we focus on the $1^+$
channel at LO.
To calculate the recombination rate $K_3(p_E)$, the absolute square
of the matrix element for the recombination process
in the $J^P=1^+$ channel, $\mathcal{M}^{1m_J}$ ($m_J\in\{-1,\,0,\,1\}$),
 has to be integrated over incoming momenta
$\bs{p}_1,\,\bs{p}_2,\,\bs{p}_3$, summed over $m_J'$ and divided by
the three-body phase space $\phi_3(p_E)$, 
{\it i.e.},
\begin{align}
  \label{eq:recom-exp}
	\nonumber
	K_3(p_E)
	=&\ \frac{1}{\phi_3(p_E)}
	\left(\prod_{i=1}^3\momint{3}{p_i}\right)
	(2\pi)^4\,
	\delta^{(3)}\left(\sum_{i=1}^3\bs{p}_i\right)
	\delta\left(\frac{p_E^2}{m}-\sum_{i=1}^3\frac{p_i^2}{2m}\right)
	\\
	&\quad\times\sum_{m_J'}\,
	\left|
		i\mathcal{M}^{1m_J}\left(\{\bs{p}_i\};\,p_E\right)
	\right|^2\,,
\end{align}
where $p_E^2/m\equiv E$.
This expression can partially be evaluated analytically by integrating over
the $\delta$-functions as will be discussed below.

Data for three-body recombination is only available at finite
temperature $T\sim\unit{\mu K}$~\cite{Waseem:2018}. For this
reason, we have to calculate the thermal average of $K_3$. The energy
$E$ is distributed according to the Boltzmann distribution of three
equal-mass particles which is proportional to $E^2 \exp[-E/(\kB T)]$;
see, {\it e.g.}, Refs.~\cite{PhysRevLett.90.053202,
  PhysRevLett.93.123201}. In terms of $p_E$, it follows that
\begin{equation}
	\label{Eq:K3T}
	\left\langle K_3\right\rangle(T)
	=\frac{1}{(m\,\kB T)^3}
	\int_0^\infty\text{d}p_E\ 
	p_E^5\,e^{-p_E^2/(m\,k_\text{B} T)}
	K_3(p_E)\,.
\end{equation}
The thermal average is directly proportional to the experimentally
measured loss coefficient~\cite{PhysRevLett.83.1751,
  Waseem:2018}
\begin{equation}
  \label{Eq:L3T}
	L_3(T)=\frac{3}{6}\left\langle K_3\right\rangle(T)\,.
\end{equation}

We reiterate that our focus is on three-body
recombination into deep dimers. However,  the same formalism can be
used to study recombination into shallow dimers, {\it i.e.}, three-body
recombination for positive two-body scattering volume $a_1$. A preliminary
study of this case can be found in Ref.~\cite{tuprints8778}.

\section{Three-body recombination matrix element}
\label{sec:diat-atom-ampl}
In order to calculate the matrix element $\mathcal{M}^{1m_J}$
in Eq.~\eqref{eq:recom-exp}, we require the atom-diatom
scattering amplitude $T^\onepl$ in off-shell kinematics.
This amplitude can be obtained by solving an integral equation
derived from effective field theory. In general, the total angular
momentum $\bs{J}=\bs{l}+\bs{L}$ is the sum of the atom-atom orbital
angular momentum $\bs{l}$ ($l=1$) and the diatom-atom orbital angular
momentum $\bs{L}$, implying $P=(-1)^{1+L}$. Thus, at small energies,
the leading partial wave channel $J^P=1^+$ is given by the single
combination $l=L=J=1$. We restrict
ourselves to this channel at LO. The theoretical uncertainty
introduced by this approximation will be addressed at the end of
this chapter. The partial-wave projected integral equation for the
amplitude $T^\onepl$ reads \cite{tuprints8778}
\begin{align}
	\nonumber
	T^\onepl&\args{p}{p'}{p_E}
	=-V^\onepl\args{p}{p'}{p_E}
	\\
	&+
	\int_0^\Lambda \frac{\d q\,q^2}{2\pi^2}\,
	V^\onepl\args{p}{q}{p_E}\,
	G_\pi\left(\tilde{k}(q^2)
        \right)
	T^\onepl\args{q}{p'}{p_E}~,
	\label{Eq:Faddeev}
\end{align}
where
\begin{equation}
  \tilde{k}(q^2) = i\sqrt{-p_E^2+\frac{3}{4}q^2-i\epsilon}\,,
  \label{eq:kmom}
  \end{equation}
is a momentum variable and 
$G_\pi$ is the Breit-Wigner-like LO diatom propagator
\begin{equation}
	\label{Eq:Gpi}
  iG_\pi(\tilde{k})=i\left[
    -a_1^{-1}(B)+\frac{r_1}{2}\tilde{k}^2
    -i\kres^3(B)\,\theta(B-B_0)\right]^{-1}~,
\end{equation}
which contains information on the two-body effective range parameters.
Moreover, $p$ ($p'$) denotes the incoming
(outgoing) atom-diatom relative momentum and
$p_E\equiv i(-mE-i\epsilon)^{1/2}$ is the
momentum scale set by the total kinetic energy.
The quantity
\begin{align}
  \label{eq:exchange_potential}
	V^\onepl \args{p}{q}{p_E} =\ &8\pi \left[ Q_0-Q_2\right]
	\left(
		\frac{p_E^2-p^2-q^2}{pq}
	\right)
	+ \frac{H^\onepl(\Lambda) pq}{\Lambda^2}
\end{align}
is the exchange potential arising from partial wave projection of the
single-atom exchange contribution where $Q_0$ and $Q_2$ are
Legendre functions of the second kind in the convention of
Ref.~\cite{AbramowitzStegun}.
The integral equation~\eqref{Eq:Faddeev} is equipped with a
momentum cutoff of natural order, $\Lambda\sim r_1$ or larger, which
also appears in the three-body force term. The three-body force
$H^\onepl(\Lambda)$ is required for renormalization of the atom-dimer
amplitude~\cite{Braaten:2011vf} and depends on the cutoff $\Lambda$.
Details of the derivation including the Feynman rules and partial wave
projection can be found in Appendices~\ref{App:L3} and
\ref{App:Faddeev}.
Note that for $B>B_0$, the diatom is unstable and does not
represent an asymptotic state. However, the quantity
$T^\onepl\args{p}{p'}{p_E}$ is required in off-shell kinematics
to calculate the three-body
recombination matrix element $\mathcal{M}^{1m_J}$.

\begin{figure}
	\includegraphics[scale=1]{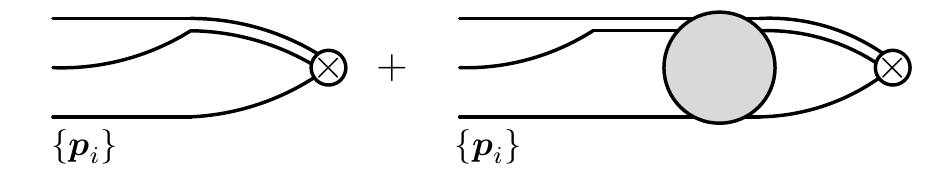}
	\vspace{-0.5cm}
	\caption{\label{Fig:Recombination}Matrix element for three-body
          recombination into a deeply bound state. The cross indicates the
          pointlike vertex for recombination into the deep dimer.
          The gray blob denotes the atom-diatom amplitude $T^\onepl$ which is
          the solution of Eq.~\eqref{Eq:Faddeev}.}
\end{figure}
The three-body recombination matrix element $\mathcal{M}^{1m_J}$,
depicted in Fig.~\ref{Fig:Recombination},
depends on the three incoming atom momenta $\bs{p}_i$ ($i\in\{1,\,2,\,3\}$).
For identical fermions, the matrix element must be antisymmetric under
exchange of each pair of momenta $\bs{p}_i,\bs{p}_j$ ($i\ne j$). This property
is automatically taken care of by the anticommutation relations of the atom
field operators $\psi,\psi^\dagger$ introduced in Appendix~\ref{App:L3}.
Applying the Feynman rules dictated
by the effective Lagrangian of Eq.~\eqref{Eq:L2} in the Appendix
to the two diagrams in Fig.~\ref{Fig:Recombination}, we find
\begin{align}
  \label{eq:matrix-element-deep}
  \nonumber
  i\mathcal{M}^{1m_J}
  \left(\{\bs{p}_i\};\,p_E\right)
  =&\ iF^\onepl(\Lambda) \,3\,
     \sqrt{\frac{2\pi}{m}}\sum_{\pi\in\, \mathcal{C}_3}
     \tens{\bs{p}_{\pi(1)}-\bs{p}_{\pi(2)}}{1m_l}\,
     G_\pi
     \left(
        \tilde{k}(p_{\pi(3)}^2)
     \right)
  \\
  &\hspace{-4cm}\times
  \Bigg(
     \Cl{1m_L}{1m_l}{1m_J}\tens{\bs{p}_{\pi(3)}}{1m_L}
  -\momint{3}{q}\,
     T^{m_l,m_l'}\!\args{\bs{p}_{\pi(3)}}{\bs{q}}{p_E}
     G_\pi\left(
        \tilde{k}(q^2)
     \right)
     \Cl{1m_L}{1m_l'}{1 m_J}\tens{\bs{q}}{1m_L}
     \Bigg)\,,
\end{align}
where 
$\Cl{L\,m_L}{l\,m_l'}{J\,m_J}$ is a Clebsch-Gordan coefficient that
couples the angular momenta $\bs{l}$ and $\bs{L}$ to $\bs{J}$ and
the sum is over all even permutations $\pi$ of $(1\,2\,3)$, denoted
$\mathcal{C}_3$. Sums over $m_l$, $m_l'$, and $m_L$ are implicit.
The coefficient of the pointlike vertex for recombination into the
deep $p$-wave dimer is given by $F^\onepl(\Lambda)$. This
regulator-dependent constant acts as
a short-range optical potential. The general equation
for $T^{m_l,m_l'}$ including all partial waves is given in Appendix
\ref{App:Faddeev}.
At LO, the unprojected amplitude $T^{m_l,m_l'}$ in
Eq.~\eqref{eq:matrix-element-deep} reduces to its $1^+$ component, {\it i.e.},
\begin{equation}
	T^{m_l,m_l'}\!\args{\bs{p}}{\bs{q}}{p_E}
	= T^\onepl\!\args{p}{q}{p_E}
	4\pi\,\sum_{m_J}
	\left(
		\bs{Y}_{(1,1)1m_J}(\hat{\bs{p}})
	\right)^{m_l}
	\left(
		\bs{Y}_{(1,1)1m_J}(\hat{\bs{q}})
	\right)^{m_l'\ \ast}
\end{equation}
in the convention of Eq.~\eqref{Eq:VPWDecomposition}.
The tensor structure $\tens{\cdot}{1m}$ in Eq.~\eqref{eq:matrix-element-deep} is defined in Eq.~\eqref{Eq:tensorstructure}.

To evaluate the expression for the recombination rate in
Eq.~\eqref{eq:recom-exp}, we further need the three-body phase space
\begin{equation}
	\label{Eq:PhaseSpace}
	\phi_3(p_E)
	=\frac{m\,p_E^4}{24\sqrt{3}\,\pi^2}\,.
\end{equation}
Inserting Eq.~\eqref{eq:matrix-element-deep} and integrating over
the $\delta$-functions, we obtain
\begin{align}
	\nonumber
	K_3(p_E)=\ &
	\frac{|F^\onepl(\Lambda)|^2}{m}\,\frac{432\sqrt{3}}{p_E^4}
	\int_0^{\frac{2}{\sqrt{3}}p_E}\!\!\!\text{d}p_A\,p_A
	\int_0^{\frac{2}{\sqrt{3}}p_E}\!\!\!\text{d}p_B\,p_B
	\ \theta(1-|x_0|)
	\\
	\label{Eq:K3}
	&\times \Big[
		\left|
			p_A\,J(p_B;\,p_E)
		\right|^2
		+ 2\,\text{Re}\,\Big(
			p_B\,J(p_A;\,p_E)
			\left[p_A\,J(p_B;\,p_E)\right]^\ast
		\Big)
	\Big]\,,
\end{align}
where
\begin{subequations}
\begin{align}
	x_0\equiv\ &\frac{1}{p_A p_B}\left(
		p_E^2-p_A^2-p_B^2
	\right),
	\\
	J(p;\,p_E)\equiv\ &G_\pi\left(
        \tilde{k}(p^2)
	\right)
         \left(
		p-\int_0^\Lambda \frac{\d q\,q^2}{2\pi^2}\,
		T^\onepl\!\left(p,\,q;\,p_E\right)
		G_\pi\left(
                \tilde{k}(q^2)
		\right)q
	\right).
\end{align}
\end{subequations}
The integral contained in the definition in $J$ diverges as
$\Lambda\rightarrow\infty$. This cutoff dependence is absorbed by
an appropriate running of 
the short-range factor $|F^\onepl(\Lambda)|^2$ with $\Lambda$.
The running of $F^\onepl$ can be obtained by making the three-body force
$H^\onepl$ in Eq.~\eqref{eq:exchange_potential} required to renormalize
$T^\onepl$ complex. Thus, the running of $F^\onepl$
with the cutoff $\Lambda$ is fully determined by the running of $H^\onepl$.
A similar procedure was previously used
to describe three-body recombination into deep  $s$-wave dimers
\cite{Braaten:2001hf}.

Before we go on, we expand on the expected size of omitted partial
waves as compared to $J^P=1^+$. For sufficiently small energies, their
contributions to the recombination rate involve at least one more
factor $E$ \cite{PhysRevA.65.010705}. Naively, one would compare this
factor to the breakdown scale $r_1^2/m$ set by the $p$-wave
effective range. For $p_E\lesssim \kres(B)=\chi_2(B)r_1$, that would yield an
\textit{a~priori} uncertainty $p_E^2/r_1^2\lesssim \chi_2^2(B)$ which is
very small ($\unit[\lesssim 0.01]{\%}$ for $B-B_0\leq \unit[0.5]{G}$).
In obtaining this estimate, we have used the threshold
laws of Ref.~\cite{PhysRevA.65.010705}. The work of Suno \etal, however,
suggests that the threshold laws fail at $p_E=\kres(B)$ \cite{Suno_2003}. At
this point, formerly subleading channels may become comparable to
$J^P=1^+$. As a consequence, we apply our LO framework only to the
low-energy region $p_E<\kres(B)$. The \textit{a~priori} LO uncertainty
at fixed energy can then be written as
$\chi_3(B,\,E)\equiv p_E^2/\kres^2(B)< 1$.

To estimate the LO uncertainty $\tilde{\chi}_3(B,\,T)$ at finite temperature, we set $p_E$ to the maximum of the Boltzmann weighting factor in Eq.~\eqref{Eq:K3T}, {\it i.e.}, to
\begin{equation}
  p_T\equiv \sqrt{\frac{5}{2}\,m\kB T}\,.
  \label{eq:def-pt}
\end{equation}
This yields the expression
\begin{equation}
	\label{Eq:chi3}
	\tilde{\chi}_3(B,\,T)\equiv p_T^2/\kres^2(B)
	\approx \frac{r_1}{2}\,\ResFactor\,\frac{5}{2}\,m\kB\,\frac{T}{B-B_0}\,,
\end{equation}
where Eqs.~\eqref{Eq:Feshbach} and \eqref{eq:defkres} have been used.
Note that we aim to determine $r_1$ from the data and thus do not
use the approximate relation~\eqref{Eq:FermireValue}.

\section{Results}
\label{Sec:Results}
\subsection{Fit of free parameters and comparison with experiment}
\begin{figure}
	\includegraphics[scale=1]{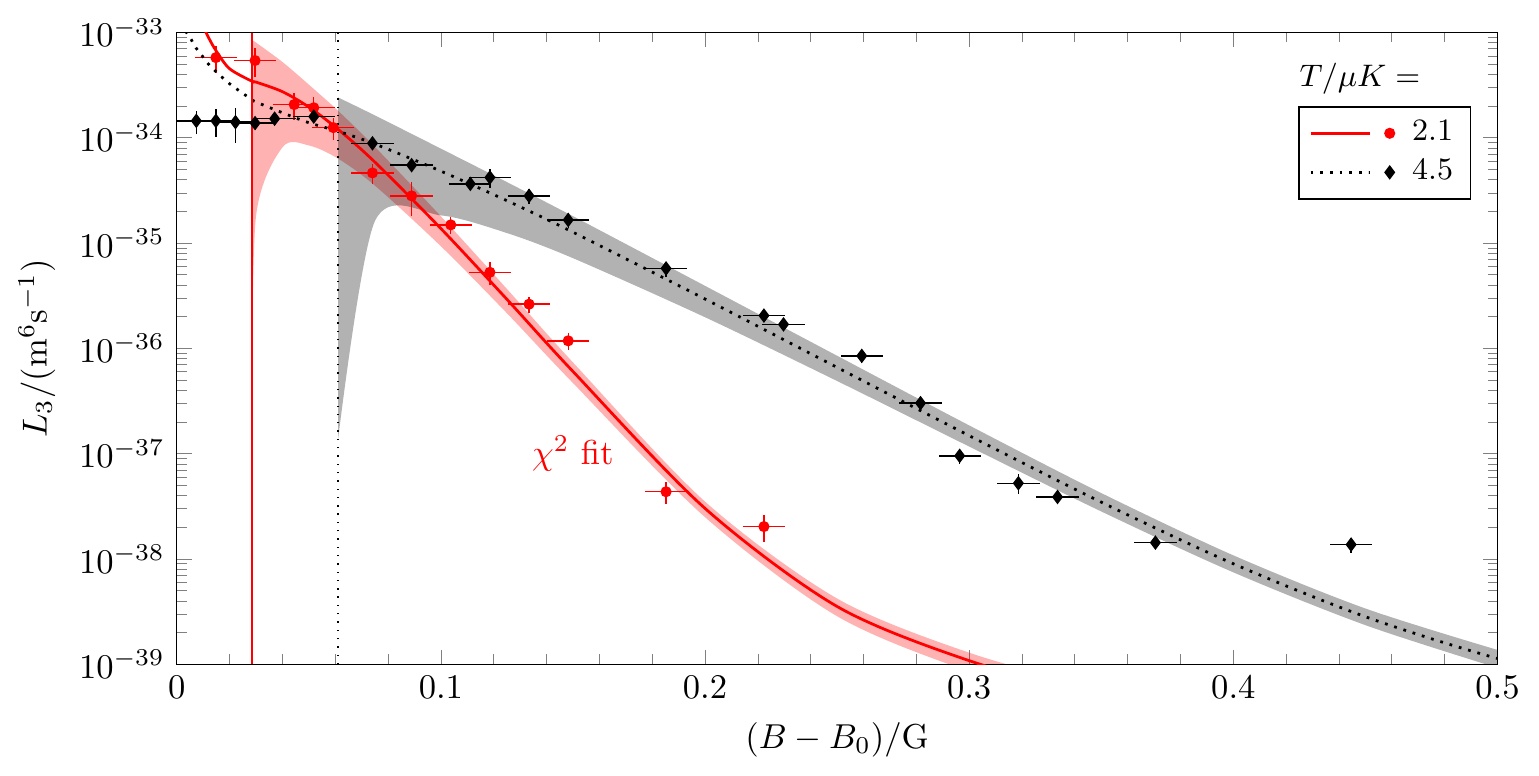}
	\caption{\label{Fig:L3(B)}Loss coefficient $L_3$ as function
          of the field detuning $B-B_0$ for two temperatures $T$ at
          $\Lambda =0.4\,\aB^{-1}$. Vertical gridlines mark points
          at which $\kres(B)=p_T$. They separate the respective unitary
          (left) from the nonunitary (right)
          regime. The solid red curve is a $\chi^2$ fit in
          $(r_1,\,H^\onepl,\,F^\onepl)$ to the $T=\unit[2.1]{\mu K}$
          data in the nonunitary regime for
          $\ResFactor=\unit[-2.8\times 10^6\,\aB^3]{G}$. The black
          dotted curve is the resulting prediction for $T=\unit[4.5]{\mu
            K}$. Uncertainty bands in the nonunitary regime follow
          from NLO corrections of order $p_T^2/\kres^2(B)< 1$ and from
          the experimental uncertainty of $\ResFactor$. Naive
            application of the theory to the unitary regime leads to
            an overestimation of the data.}
\end{figure}
The data obtained in Ref.~\cite{Waseem:2018} can be divided
into two different regimes, the unitary regime and the non-unitary
regime. We define the unitary regime as the temperature domain in
which the resonance momentum $k_{\rm res}$ is smaller than the thermal
momentum scale $p_T$ defined in Eq.~\eqref{eq:def-pt}.
For a given resonance momentum $k_{\rm res}$, the unitary regime sets in
at temperatures larger than
\begin{equation}
  \label{eq:unitary}
  T_{\rm unitary} > \frac{2 k_{\rm res}^2}{ 5 m k_B}~.
\end{equation}
We do not expect our EFT to work in this regime since
the expansion parameter $\tilde{\chi}_3(B,\,T)\gtrsim 1$. 
For convenience, we will also drop the superscript $1+$
in the three-body terms $H$ and $F$ from now on.

We use our approach to fit the effective range $r_1$, the three-body
force $H$, and the short-distance three-body parameter $F$ to the
experimental data for $T = 2.1~\mu$K from Ref.~\cite{Waseem:2018}.
Our results are renormalization group invariant and independent of
$\Lambda$, but for definiteness we use an ultraviolet cutoff
$\Lambda = 0.4~a_B^{-1}$ in the integral equations.  We find an
effective range $r_1=-0.11(2)\, a_B^{-1}$ which is of the same order
of magnitude as the result by Waseem \etal\ but deviates by 60 \%. For
the three-body force, we find $H = 4\substack{+5\\-7}$ and for the
short-distance three-body parameter we obtain
$\log_{10}\left(F/m^2\right) = 49.8\substack{+0.2\\-0.1}$. We
emphasize that these two quantities are not observables and
depend on the ultraviolet cutoff $\Lambda$
used to solve Eq.~\eqref{Eq:Faddeev}.

The red solid line in Fig.~\ref{Fig:L3(B)} shows the fitted loss
coefficient $L_3$ in comparison to the experimental data (red
circles). Experimental data in the unitary regime was excluded in the
fit. These data points are to the left of the red solid line in that
figure.  Once the parameters of our approach are determined, we can
use Eqs.~\eqref{Eq:K3T} and \eqref{Eq:L3T} to predict the loss rate at
a different temperature. Figure~\ref{Fig:L3(B)} shows also the
resulting prediction
for the loss rate at a temperature 4.5~$\mu$K (black dotted line) in
comparison to the experimental data from
Ref.~\cite{Waseem:2018} (black diamonds).  Our results describe
data at these small temperatures relatively well.

\begin{figure}
  \includegraphics[scale=1]{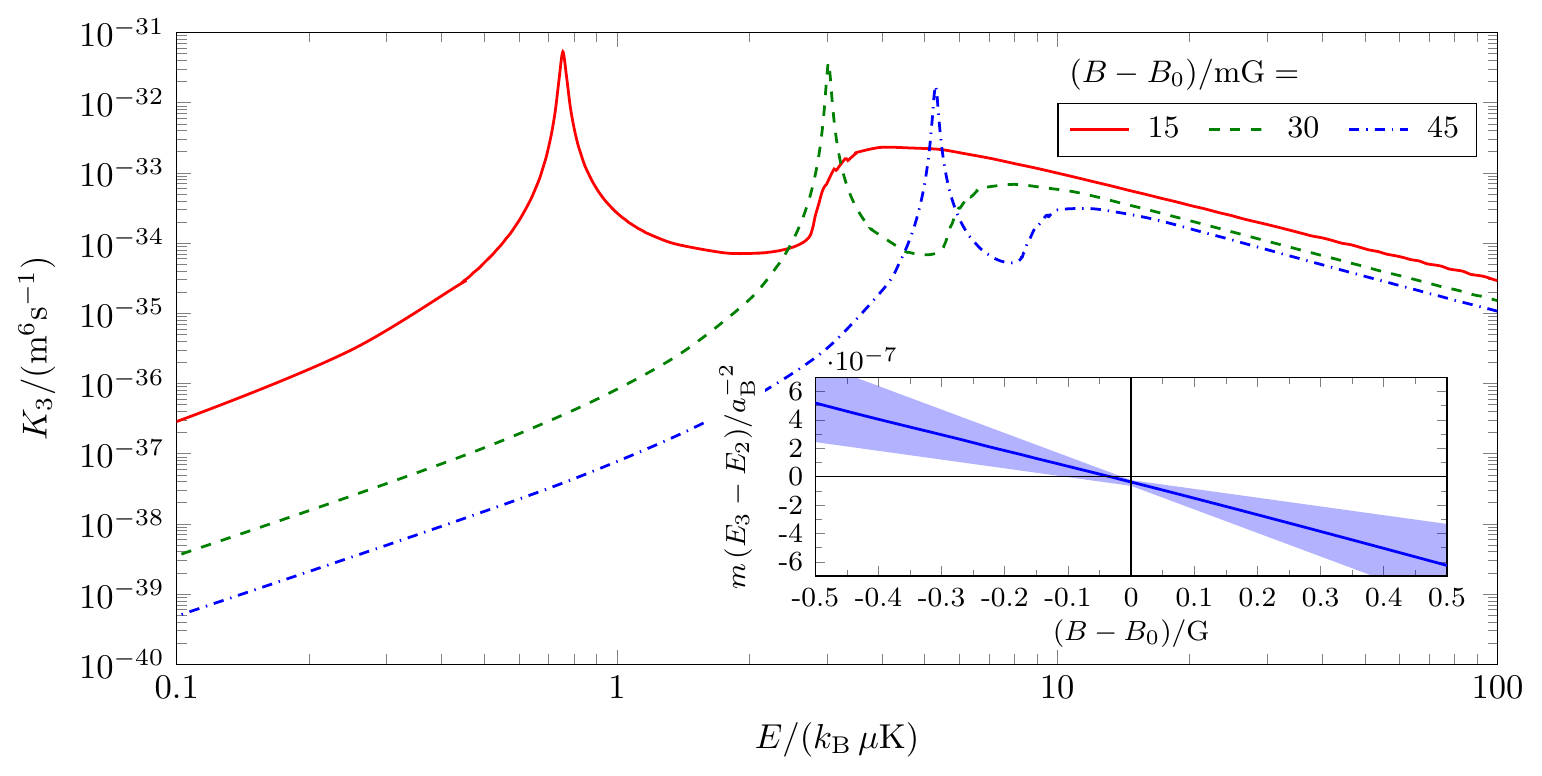}
  \caption{\label{Fig:K3(E)}Recombination rate coefficient $K_3$ as function of
    the energy $E$ for several field detunings $B-B_0$ at
    $\Lambda =0.4\,\aB^{-1}$.
    The inset shows the rescaled energy gap $m(E_3-E_2)$ as a function
    $B-B_0=0$ with an error estimate given by the shaded band.
  }
\end{figure}
The finite-temperature averaging of the energy-dependent recombination
rate coefficient $K_3$ 
smears out the resonance features from the three-body loss
rate. In Fig.~\ref{Fig:K3(E)}, we show $K_3$ as
a function of the energy for different magnetic field detunings.
The curves have been
generated with the parameters obtained in the fit to the
$T = 2.1~\mu$K data from Ref.~\cite{Waseem:2018} discussed above.
They
display a strong peak at lower energies followed by a sudden
increase and a smooth fall off. The peak on the
left is caused by the existence of a three-body resonance below the
two-body resonance. Its position is controlled by the three-body force
$H$. The sudden rise in $K_3$ to the right of the peak
is the signature of the two-body resonance.

Given the fit results for $r_1$ and $H$, the difference of the
three- and two-body resonance energies, $E_3-E_2$, is a function
of the scattering volume, {\it i.e.}, the magnetic field detuning, only.
Thus, our approach allows us to predict the magnetic field dependence of the
three-body resonance energy $E_3$. In the inset of Fig.~\ref{Fig:K3(E)},
we show the rescaled energy gap $m(E_3-E_2)$ as a blue line.
For $B-B_0>0$, the three-body resonance is below the two-body resonance
and $E_3-E_2$ is negative. Near $B-B_0=0$,
$E_3$ is linear in $B$ just like $E_2$ and the three-body
resonance crosses the two-body energy.
In the region $B-B_0<0$, the two-body resonance turns into a shallow
bound state and the energy difference $E_3-E_2$ becomes positive.

Note that corrections to our prediction should arise from omitted $J^P$
channels. Contributions of these channels to the loss rate
coefficient $L_3(B)$ in the
fit regime should have relative sizes $0\leq \chi_3(B,T)\leq 1$;
see Eq.~\eqref{Eq:chi3}. Gaussian uncertainty propagation then implies a
relative uncertainty of $\unit[46]{\%}$ for the $\chi^2$ value of
our fit. This number
can be used as an estimate for the uncertainties of the offset $\alpha$ and
slope $\beta$ of the $m(E_3-E_2)=\alpha + \beta (B-B_0)$ curve. We obtain
$\alpha=-0.4(2)\times 10^{-7}\,\aB^{-2}$ and
$\beta=-11(5)\,\times 10^{-7}\,\aB^{-2}/G$ which yields the blue band
depicted in the inset of Fig.~\ref{Fig:K3(E)}.

\begin{figure}
	\includegraphics[scale=1]{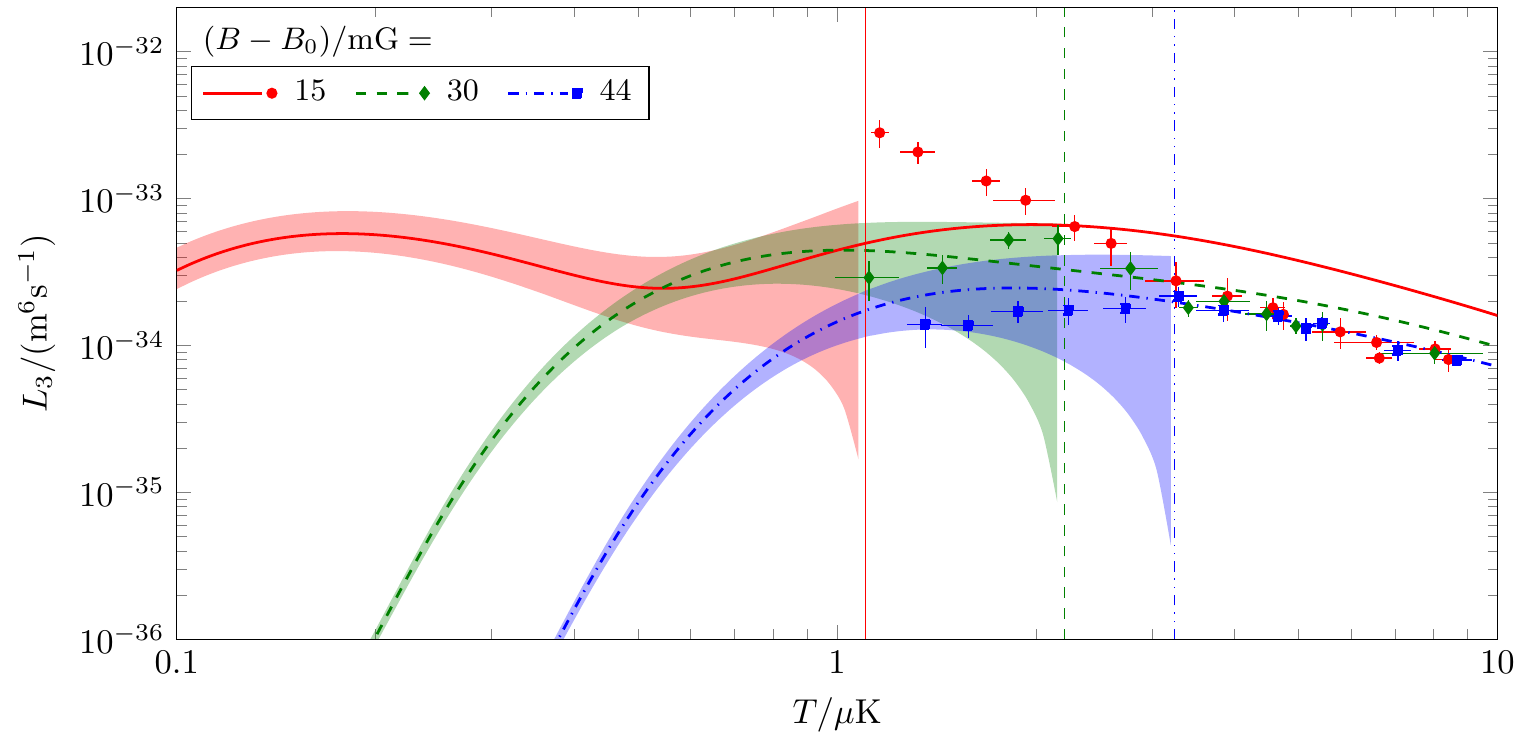}
	\caption{\label{Fig:L3(T)}Loss coefficient $L_3$ as function
          of the temperature $T$ for several field detunings $B-B_0$
          at $\Lambda =0.4\,\aB^{-1}$ ($\chi^2$-fit
          predictions). Vertical gridlines at $\kT=\kres$ separate the
          nonunitary (left) from the unitary (right)
          regime. Uncertainty bands are obtained as in
          Fig.~\ref{Fig:L3(B)}. For clarity, they are omitted for the
          higher detunings.}
\end{figure}

Finally, Fig.~\ref{Fig:L3(T)} shows the temperature averaged
loss rate $L_3$ as a function of the temperature $T$ for
different detunings $B-B_0$. The vertical lines denote the beginning of the
unitary regime as given by Eq.~\eqref{eq:unitary}. 
While they reproduce the  expected $T^{-2}$ behavior at large temperatures,
the curves are not independent of $B$ in the unitary
regime. Instead, they are separated by factors of 1.3-1.5.
We expect that effects from the unitary cut will be important in
this region where the system is able to probe the resonance
peak~\cite{Bedaque:2003wa}.
Moreover, the contribution from  spin-parity channels
different from $J^P = 1^+$ could be important.
Even though suppressed close to $E = 0$,
they might contribute significantly at finite
temperature. It would also be instructive to iterate effects of the
short-range three-body factor $F$.
A nonperturbative treatment would presumably change the
behavior at larger energies.
Understanding the loss rate in this region is left to future work.

\subsection{Comparison with Waseem \etal}

\begin{figure}
  \includegraphics[scale=1]{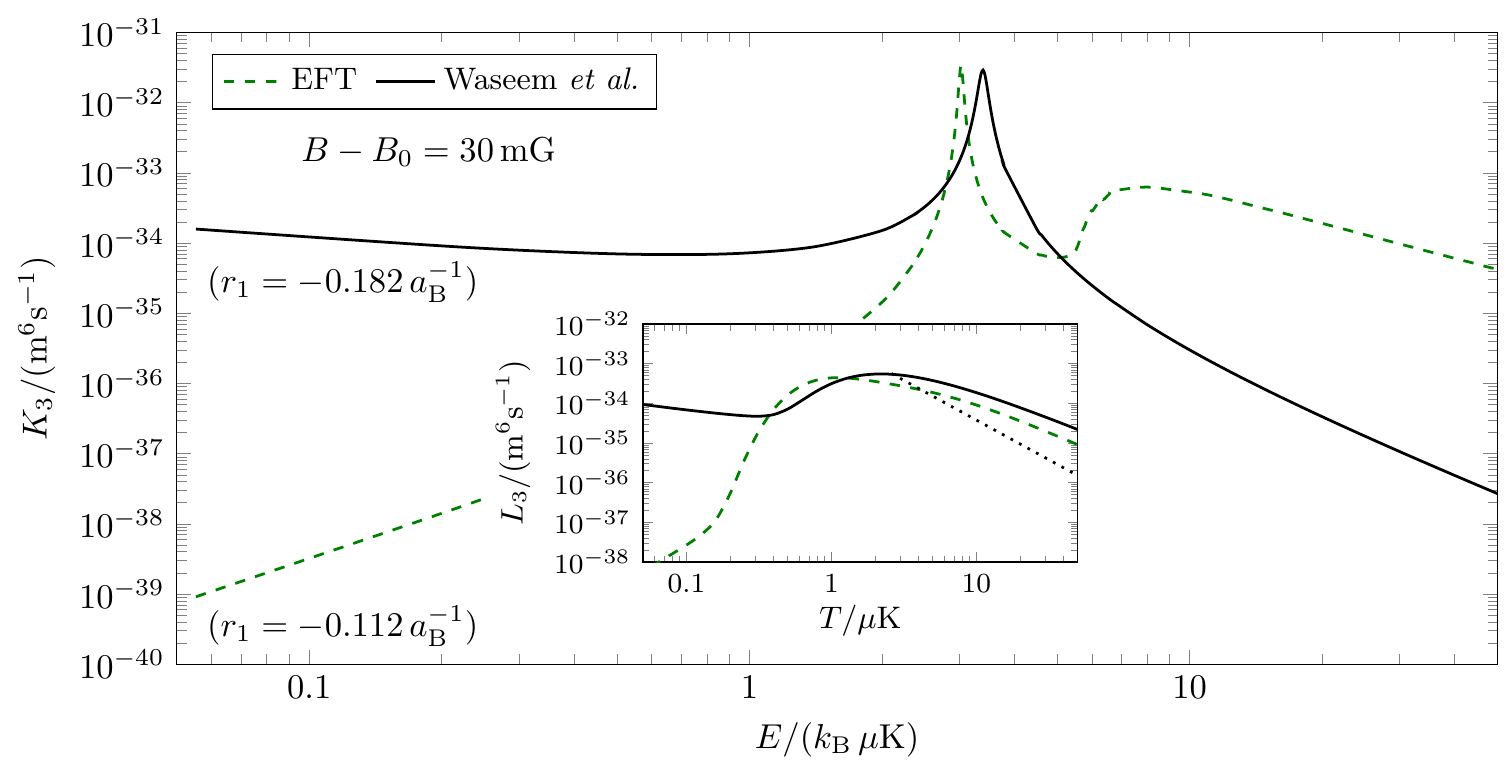}
  \caption{\label{Fig:K3comp}Comparison of the energy-dependent
    recombination rate coefficient $K_3$ from our EFT calculation (dashed curve)
    and from the model by Waseem \etal~\cite{Waseem:2018} (solid
    curve) as function of the energy $E$ for a magnetic field detuning
    of $B-B_0 = \unit[30]{mG}$.  The inset shows the same comparison
    for the temperature-dependent loss coefficient $L_3$ where the
    dotted curve shows the $E^{-2}$ behavior expected from
    unitarity. The dramatically different behavior of the
    recombination rates at low energies leads to different
    predictions for the loss rates at temperatures below 0.5~$\mu$K
    only.}
\end{figure}
We now compare our results to the  two-body model
employed by Waseem \etal~\cite{Waseem:2018},
\begin{eqnarray}
  \label{eq:K3Waseem}
  K_3^{\rm Waseem}(E) = \frac{144\sqrt{3}\pi^2}{m^3
  E^2}\frac{\Gamma_e \Gamma_d}{(E-E_b)^2+(\Gamma_e+\Gamma_d)^4/4}~,
\end{eqnarray}
where $\Gamma_e = -4 \sqrt{m} E^{3/2}/r_1$,
$\Gamma_d=-4/(m r_1 a_1)$, and
$E_b = k_{\rm res}^2/m$ is the real part of the
resonance energy.\footnote{Note that we have converted Eq.~\eqref{eq:K3Waseem}
  to our conventions where $\hbar=1$.}
While the temperature averaged loss coefficient $L_3$ looks very
similar to our result, the energy-dependent
recombination rate coefficient $K_3$ behaves very differently.

This can be seen in Fig.~\ref{Fig:K3comp}, where we compare the three-body
recombination model in Eq.~\eqref{eq:K3Waseem} and our results for a
magnetic field detuning of $B-B_0 = 15$~mG.
While the temperature-averaged loss coefficient $L_3$ shown in the inset
for both approaches agrees for temperatures $T>1\,\mu$K  measured
in the experiment by Waseem \etal,
the energy-dependent rate coefficient $K_3$ shows stark differences.
The rate calculated in the EFT approach falls off at small energies
as $E^2$ while the model in Eq.~(\ref{eq:K3Waseem}) grows
as $E^{-1/2}$. Both approaches show a peak around $E=3\,k_B\mu$K but the
microscopic origin is very different. In the model
of Waseem \etal\ it is simply associated with the Breit-Wigner form
put in by hand, while the resonance in the
full three-body treatment in the EFT framework is dynamically
generated by the $p$-wave atom-atom interactions.
Finally, there is a shoulder around $E=7\,k_B\mu$K
in the EFT framework which is not present in the model.

Currently, the data for the temperature averaged loss rate
is not able to distinguish between the two approaches and
thus the underlying microphysics.
The inset of Fig.~\ref{Fig:K3comp}, however, suggests that a
new measurement at lower temperatures $T \leq 0.5$ $\mu$K should be
able to distinguish between the two scenarios. 

\section{Summary}
\label{sec:summary}
In this work, we have considered the temperature dependent three-body
loss rate for a gas of identical fermions with resonant $p$-wave
interactions. We have used an effective field theory approach to
derive integral equations that describe the scattering of three
fermions and used them to evaluate the rate for three-body
recombination into deeply bound dimers and another atom.  Our approach
requires the determination of four parameters to become predictive; the
scattering volume $a_1$, the effective range $r_1$, and two pieces of
three-body information to determine the short-range three-body
parameters $H$ and $F$. The latter acts as an optical potential and
parameterizes the coupling of the three-body system to final state
channels with deep dimers.  We have used the known magnetic field
dependence of the scattering volume $a_1$ and the experimental data
from Waseem \etal~\cite{Waseem:2018}, to fit the remaining parameters
$r_1$, $H$, and $F$. Once these parameters are determined, we were
also able to determine the position and magnetic field dependence of a
fermionic three-body state with $J^P = 1^{+}$. We note that this
three-body resonance could lead to interesting features in the
three-body recombination rate on the positive scattering volume side
of the Feshbach resonance, where the three-body state
is a resonance close to the atom-dimer threshold. For
positive scattering volume, it would also be interesting to consider
atom-dimer relaxation as an additional benchmark to our approach.

While the temperature averaged rate coefficient $L_3$ from the
two-body resonance model employed in \cite{Waseem:2018} and from our
three-body calculation are very similar, the energy-dependent
recombination rate $K_3(E)$ shows significant differences.  New
experiments at lower temperatures $T \leq 0.5$ $\mu$K should be able
to distinguish between the two scenarios and determine the nature of
the microscopic physics responsible for the loss processes.

In summary, we have shown that the three-body recombination rate coefficient
$K_3(E)$ for spin-polarized $^6$Li atoms with resonant $p$-wave
interactions possesses interesting features due to two- and three-body
resonances which can be seen at low temperatures. We also have demonstrated
that three-body loss data can be used to extract detailed information
on the two-body interaction once a reliable parameterization
is established and used in a full three-body
calculation. Our effective field theory approach has the additional
advantage that the interaction is directly given in terms of the
effective range parameters.

Waseem {\it et al.} recently described the unitary regime using a
model for cascade processes that leads to a modified three-body loss
coefficient $L_3$ \cite{Waseem:2019}. Our uncertainty analysis led us
to exclude this regime as more partial wave channels are expected to
contribute to the loss rate at these temperatures. It would therefore
be interesting to include additional partial wave channels in our
calculation of the total loss rate in order to test
their result and to analyze whether it is really necessary to include
additional recombination mechanisms to achieve agreement with the
data.

We finally note that the S-wave scattering between the
$\ket{F=1/2,\,m_F=+1/2}$ and $\ket{F=1/2,\,m_F=-1/2}$ hyperfine states
of $^6$Li is moderately large (and negative) at the magnetic fields
considered in this work \cite{PhysRevA.89.052715}.
The study of recombination losses in this two-component
system of identical fermions with large $s$- and $p$-wave scattering
lengths might therefore lead to interesting features such as a
competition between odd and even parity recombination channels.

\begin{acknowledgments}
  M.S. thanks the nuclear theory groups of UT Knoxville and Oak Ridge
  National Laboratory for their kind hospitality during his research
  stay.  H.W.H and L.P. thank the Institute for Nuclear Theory at the
  University of Washington for its kind hospitality and stimulating
  research environment. L.P. thanks the nuclear theory group of TU
  Darmstadt for its hospitality during the final stages of this work.
  This research was supported by the Deut\-sche
  For\-schungs\-ge\-mein\-schaft (DFG, German Research Foundation)~--
  Pro\-jekt\-num\-mer 279384907~-- SFB 1245, by the National Science
  Foundation under Grant No.  PHY-1555030, by the Office of Nuclear
  Physics, U.S.  Department of Energy under contracts
  No. DE-AC05-00OR22725 and No. DE-FG02-00ER41132.
\end{acknowledgments}

\appendix

\section{Lagrangian}
\label{App:L3}
The EFT Lagrangian for the spin-polarized fermions can be split
into three parts:
$\mathcal{L}=\mathcal{L}_1+\mathcal{L}_2+\mathcal{L}_3$.  For
$\Li$ atoms with mass $m=\unit[6.0151223(5)]{u}$
\cite{IUPACReview}, the one-body part
\begin{equation}
	\label{Eq:L1}
	\mathcal{L}_1 
	=\psi^\dagger
	\left[
		\idz+\kinterm{2m}
	\right]
	\psi
\end{equation}
can be written in terms of scalar fields $\psi,\psi^\dagger$ which
anticommute.

The remaining parts comprise all two- and three-body
contact terms compliant with Galilean symmetry. They are formulated
using bosonic fields $\pi_{m_l}, \pi^\dagger_{m_l}$
($m_l\in\{-1,\,0,\,1\}$), which annihilate and create two atoms in a
$p$-wave, respectively. At LO, $\mathcal{L}_2$ reads
\begin{align}
	\nonumber
	\mathcal{L}_2
	=&\ \pi_{m_l}^\dagger
	\left[
		\Delta
		+\left(\idz+\kinterm{4m}\right)
		+\cdots
	\right]
	\pi_{m_l}
	\\
	\label{Eq:L2}
	&\ - \frac{g}{\sqrt{2}}
	\left[
		\pi_{m_l}^\dagger\left(\psi
		\tens{-i\galder_{\!\!2}}{1m_l} \psi
		\right)
		+\Hc
	\right],
\end{align}
where ``$\Hc$'' is the Hermitian conjugate and the sums over $m_l$ are
implicit. The $p$-wave nature of the diatom manifests itself in the
tensor structure $\tens{\cdot}{1m_l}$. Together with the
Galilean-invariant derivative
$\galder_{\!\!2}\equiv(\overleftarrow{\bs{\nabla}}-\overrightarrow{\bs{\nabla}})/2$,
it contributes a factor
\begin{equation}
	\label{Eq:tensorstructure}
	\tens{\bs{k}}{1m_l}\equiv (4\pi/3)^{1/2}\,k\,Y_1^{m_l}(\hat{\bs{k}})
\end{equation}
to the $\pi$-$\psi\psi$ vertex. Here, $\bs{k}=(\bs{k}_1-\bs{k}_2)/2$
is the atom-atom relative momentum. To describe a shallow $p$-wave
resonance, two real low-energy parameters are required
\cite{Bertulani:2002sz,Bedaque:2003wa}. They are chosen as $\Delta$
and the coupling $g$ which both will be matched to observables.

The three-body part $\mathcal{L}_3$ contains three-body interactions which
eliminate potential divergences in different spin-parity channels
$J^P$. At LO, a single three-body force ($J^P=1^+$) enters the
theory. It takes the form
\begin{equation}
	\label{Eq:L3}
	\mathcal{L}_3 
	=-C_0^{\onepl} \frac{12\pi}{mg^2}
	\left(
		\psi\,\bs{\pi}
	\right)_{m_J}^{\onepl\,\dagger}
	\left(
		\psi\,\bs{\pi}
	\right)_{m_J}^\onepl
	+\cdots
\end{equation}
with the  $J^P=1^+$ ($l=L=J=1$) field combinations
\begin{equation}
	\label{Eq:FieldCombination}
	\left(
		\psi\,\bs{\pi}
	\right)_{m_J}^\onepl
	=\sqrt{3}\,\Cl{1m_L}{1m_l}{1m_J}
	\psi\tens{-i\galder_{\!\!3}}{1m_L} \pi_{m_l}
\end{equation}
and the Galileian-invariant derivative $\galder_{\!\!3}\equiv
(2\overleftarrow{\bs{\nabla}}-\overrightarrow{\bs{\nabla}})/3$.
Further, we define dimensionless three-body force $H^\onepl$ with
\begin{equation}
	C_0^\onepl(\Lambda) = H^\onepl(\Lambda)/\Lambda^2\,.
\end{equation}

\section{Faddeev equation}
\label{App:Faddeev}
The Feynman rules resulting from the Lagrangian given in the previous
section can be used to derive an integral equation for atom-dimer
scattering. For orbital angular momentum quantum numbers $l=l'=1$ and
projection $m_l$ and $m_{l'}$, respectively, this integral equation is
given by
\begin{align}
	\nonumber
	T^{m_l,\,m_l'}&\args{\bs{p}}{\bs{p}'}{p_E}
	=-V^{m_l,\,m_l'}\args{\bs{p}}{\bs{p}'}{p_E}
	\\
	+&\sum_{m_l''}\momint{3}{q}\,
	V^{m_l,\,m_l''}\args{\bs{p}}{\bs{q}}{p_E}
	\label{Eq:FaddeevFull}
        G_\pi \left(\tilde{k}(q^2)\right)
	T^{m_l'',\,m_l'}\!\args{\bs{q}}{\bs{p}'}{p_E}\,,
\end{align}
where $\tilde{k}(q^2)$ is given in Eq.~\eqref{eq:kmom}.  The particle
exchange potential is given by
\begin{subequations}
\begin{align}
	\label{Eq:VFull}
	-iV^{m_l,\,m_l'}
	&\args{\bs{p}}{\bs{q}}{p_E}
	=-i24\pi\,\frac{
		\tens{\bs{q}+\bs{p}/2}{1m_l}^\ast
		\tens{\bs{p}+\bs{q}/2}{1m_l'}
	}{p^2+q^2-p_E^2+\bs{p}\cdot \bs{q}}
	\\
	\nonumber
	=&\ -i\sum_J\sum_{L,\,L'}
	\ V^{\spec{3}{L}_J,\spec{3}{L'}_J}\!\args{p}{q}{p_E}
	\\
	\label{Eq:VPWDecomposition}
	&\quad\times
	4\pi\,\sum_{m_J}
	\left(
		\bs{Y}_{(L,1)Jm_J}(\hat{\bs{p}})
	\right)^{m_l}
	\left(
		\bs{Y}_{(L',1)Jm_J}(\hat{\bs{q}})
	\right)^{m_l'\ \ast}\,.
\end{align}
\end{subequations}

Projection onto partial waves with $l=1$, total orbital angular momentum
$L$, and total angular momentum $J$ yields 
\begin{align}
	\label{Eq:PWPot}
	\nonumber
	V^{\spec{3}{L}_J,\spec{3}{L'}_J}\!
	\args{p}{q}{p_E}
	=&\ -24\pi\,\frac{\sqrt{(2L+1)(2L'+1)}}{2J+1}
	\\\nonumber
	&\ \hspace{-2.5cm}\times \left[
		\Cl{L0}{10}{J0}\,\Cl{L'0}{10}{J0}\,\frac{1}{2}\left(
			\frac{p}{q}\,\hat{t}_{L'}+\frac{q}{p}\,\hat{t}_L
		\right)
	\right.
	\\
	\nonumber
	&\ \hspace{-1.5cm}\left.
		+pq\left(
			\frac{1}{4}\,\Cl{L0}{10}{J0}\,\Cl{L'0}{10}{J0}\,\hat{t}_J
			+(2J+1)\sum_k \Cl{L0}{10}{k0}\,\Cl{L'0}{10}{k0}\,
			\left\lbrace\begin{matrix}
				1&J&L'
				\\
				1&k&L
			\end{matrix}\right\rbrace
			\hat{t}_k
		\right)
	\right]
	\\
	&\hspace{-2.5cm}\times
	Q_{\cdot}\left(
		\frac{p_E^2-p^2-q^2}{pq}
	\right)
\end{align}
where $Q_{\cdot}$ are Legendre functions of the second kind in the convention of
Ref.~\cite{AbramowitzStegun} and 
the short notation $\hat{t}_L Q_\cdot \equiv Q_L$ was used.
For the $1^+$ channel ($l=L=J=1$), we recover Eq.~\eqref{eq:exchange_potential}.


\end{document}